\documentclass[%
reprint,
superscriptaddress,
showpacs,
 amsmath,amssymb,
prl
]{revtex4-1}

\usepackage{graphicx}
\usepackage{dcolumn}
\usepackage{bm}
\usepackage{xcolor}

\begin{document}

\title{Scalable spin-glass optical simulator}

\author{Davide Pierangeli $^\dag$}
\email{davide.pierangeli@roma1.infn.it}
\affiliation{Dipartimento di Fisica, Universit\`{a} di Roma  ``La Sapienza'', 00185 Rome, Italy}
\affiliation{Institute for Complex System, National Research Council (ISC-CNR), 00185 Rome, Italy}

\author{Mushegh Rafayelyan $^\dag$}
\affiliation{Laboratoire Kastler Brossel, \'{E}cole Normale Superi\'{e}ure-Universit\'{e} PSL, CNRS, Sorbonne Universit\'{e}, College de France, 75005 Paris, France \linebreak \normalfont $^\dag$, $^\ddag$ These authors contributed equally to this work. }

\author{Claudio Conti $^\ddag$}
\affiliation{Dipartimento di Fisica, Universit\`{a} di Roma  ``La Sapienza'', 00185 Rome, Italy}
\affiliation{Institute for Complex System, National Research Council (ISC-CNR), 00185 Rome, Italy}

\author{Sylvain Gigan $^\ddag$}
\affiliation{Laboratoire Kastler Brossel, \'{E}cole Normale Superi\'{e}ure-Universit\'{e} PSL, CNRS, Sorbonne Universit\'{e}, College de France, 75005 Paris, France \linebreak \normalfont $^\dag$, $^\ddag$ These authors contributed equally to this work. }

\begin{abstract}
\vspace*{0.2cm}
Many developments in science and engineering depend on tackling complex optimizations on large scales.
The challenge motivates intense search for specific computing hardware that takes advantage from quantum features, nonlinear dynamics, or photonics. 
A paradigmatic optimization problem is finding low-energy states in classical spin systems with fully-random interactions.
To date no alternative computing platform can address such spin-glass problems on a large scale.
Here we propose and realize an optical scalable spin-glass simulator based on spatial light modulation and multiple light scattering. 
By tailoring optical transmission through a disordered medium, we optically accelerate the computation of the ground state of large spin networks
with all-to-all random couplings. Scaling of the operation time with the problem size demonstrates optical advantage over conventional computing.
Our results point out optical vector-matrix multiplication as a tool for spin-glass problems
and provide a general route towards large-scale computing that exploits speed, parallelism and coherence of light. 
\end{abstract}

\maketitle

\section{INTRODUCTION}

Non-deterministic polynomial-time (NP) problems are crucial from biochemistry to quantum physics.
Their solution using polynomial resources requires non-deterministic Turing machines, which are unconventional computing models where a defined state can result in different outcomes \cite{Garey1979}. Alternative computing architectures exploit quantum annealing~\cite{Johnson2011, Boixo2014}, 
stochastic elements~\cite{Datta2019}, nonlinear dynamics with gain and losses~\cite{Marandi2014, McMahon2016, Inagaki2016, Vandersande2019, Inagaki2016_2, Takesue2020}, in-memory operations~\cite{Traversa2015, Cai2019}, or photon's speed and coherence~\cite{Pierangeli2019, Prabhu2020, Wu2014, Jin2020,
Farhat1985, Shen2017, Engheta2019, Brunner2018, Hamerly2019, Marcucci2020, Rafayelyan2020}.
Among them, Ising machines are special-purpose processors designed for finding a ground state of a Ising spin model. 
They are currently attracting broad attention, since tasks such as partitioning, routing, and encrypting can be mapped on Ising Hamiltonians \cite{Lucas2014}. Devices based on various physical mechanisms have been recently realized using superconducting networks \cite{Harris2018, Hamerly2019_2}, optical parametric oscillators \cite{McMahon2016, Inagaki2016, Vandersande2019}, polariton condensates \cite{Berloff2017, Kalinin2020}, coupled laser cavities \cite{Khajavikhan2020, Davidson2019}, nanophotonic circuits \cite{Prabhu2020, Gaeta2020} and spatial light modulators (SLM) \cite{Pierangeli2019, Pierangeli2020, Kumar2020, Fang2020, Pierangeli2020_2}.
Scalability with respect to the problem size is the main factor hampering their near-term application.
In fact, several Ising machines, such as the D-Wave quantum annealer \cite{Hamerly2019_2}, rely on local interactions between their elementary units, 
a fact that strongly limits long-range connectivity and imposes redundant schemes difficult to scale in practice~\cite{Lechner2015}. 
Other platforms, such as coherent Ising machines (CIMs) \cite{McMahon2016, Inagaki2016, Vandersande2019, Inagaki2016_2}, provide all-to-all connectivity and can host dense spin networks made of thousands of elements, but with couplings that are not fully programmable and assume only a few possible values.
For these reasons, the relevant NP problems that can be implemented and solved heuristically on Ising machines on large scale are still not exhaustive. 

In this Article, we report a pivotal step toward ``Ising computing" by realizing a scalable photonic device that can simulate large-scale spin problems with continuous random couplings.
We demonstrate use of optical random vector-matrix multiplications to implement the energy function of a spin-glass system.
Since the optical setting enables simultaneous processing of all spin interactions in parallel, 
our approach exhibits an optical advantage at large scale over digital computing.
The photonic hardware accelerates the solution of the spin-glass problem independently of the used algorithm,  
which suggests that our setup may potentially speed-up any minimization approach.
We apply the optical simulator to the number-partitioning problem, 
thus proving it can be useful for a vast class of practical combinatorial optimization tasks.
Although we program only the coupling distribution, our scheme may develop into a fully-programmable special-purpose optical processor
using reconfigurable transmissive elements \cite{Leedumrongwatthanakun2020, Popoff2019, Lvovsky2020}.

\begin{figure*}[t!]
\centering
\vspace*{-0.4cm}
\hspace*{-0.3cm} 
\includegraphics[width=1.80\columnwidth]{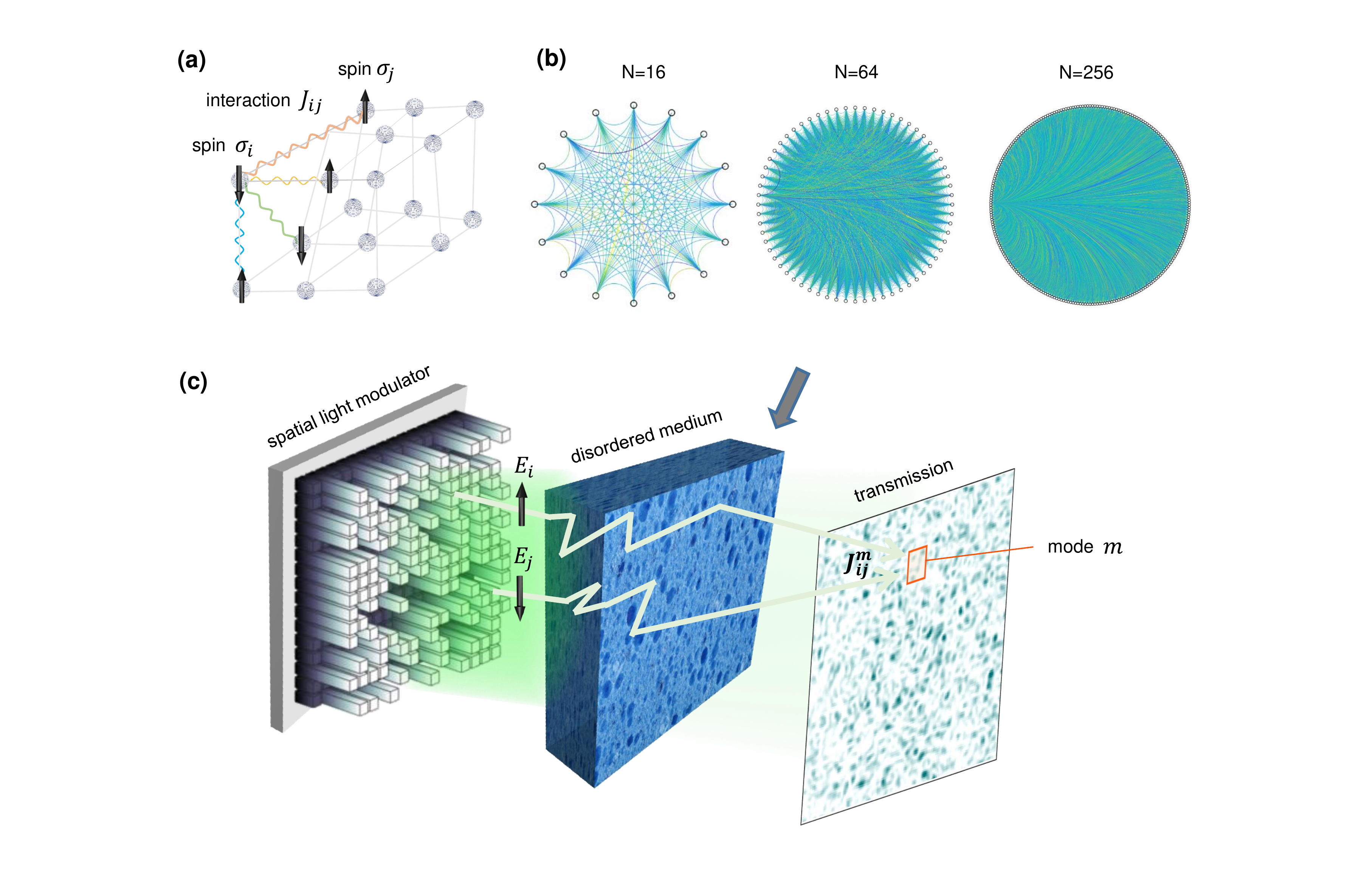} 
\vspace*{-0.4cm}
\caption{{\bf Scheme of the optical spin-glass simulator.} (a) Sketch of Ising spins on a disordered lattice. (b) Graph representation of a SG problem for various sizes. Each spin is a node of a fully-connected network where the coupling matrix $J_{ij}$ is represented by color-coded links. 
(c) Optical scheme mapping the SG model. A spatial light modulator (SLM) inscribes Ising spins in separated spatial points of the optical field $E_i$. 
The spin network is encoded in a disordered medium that mixes all the incoming modes according to its scattering matrix. 
Any $i$-th and $j$-th two spins contribute to the $m$-th output mode by a coupling coefficient~$J^{m}_{ij}$.
The computation works by optimizing the total intensity transmitted on a set of output modes.
}
\vspace{-0.2cm}
\label{Figure1}
\end{figure*}

\section{Model of the optical spin-glass simulator}

Finding the minimum energy configuration of a spin glass (SG) is a benchmark NP-hard problem \cite{Nishimori2001, Barahona1982}, 
and its computer intractability continuously inspires novel heuristic algoritms \cite{Goto2019, Roques-Carmes2020}. 
The system can be illustrated as in Fig.~1(a), where a set of $N$ unitary Ising spins $\sigma_i \in\lbrace+1,-1\rbrace$ occupies the sites of a disordered lattice. Due to strong lattice distortions, the effective interaction $J_{ij}$ between the $i$-th and $j$-th spin takes a broad spectrum of values. The quadratic SG Hamiltonian has the form 
\begin{equation}
H(\sigma)= - \frac{1}{2}\sum_{i,j=1}^{N} J_{ij} \sigma_i \sigma_j
\end{equation}
where the  $J_{ij}$ elements come from a Gaussian distribution function $P(J_{ij})$. The model is a cornerstone of statistical mechanics also known as Sherrington-Kirkpatrick (SK) model \cite{Nishimori2001}. Each problem instance corresponds to a graph of $N$ all-to-all connected nodes with a set of randomly weighted links [Fig. 1(b)]. 

The operating principle of our optical SG simulator is shown in Fig. 1(c). The basic idea is to encode the 
spins on a coherent wavefront by spatial light modulation \cite{Pierangeli2019} and their interaction on the optical transmission matrix (TM) of a disordered medium \cite{Popoff2019, Popoff2010}. 
A similar approach has been recently investigated as an instruments to access spin-glass dynamics and its complexity \cite{Leonetti2020}.
Specifically, we consider the optical field transmitted via multiple scattering $E_m = \sum_i t^m_{i} E_i$, 
where $1<i<N$ and $t^m_{i}$ is the complex TM element connecting the $i$-th input mode (spin) generated by an SLM to the $m$-th output mode
detected by a camera \cite{Popoff2010}. 
The total intensity transmitted over $M$ output modes is thus $I_{T}= \sum_m \left| E_m \right|^2 = \sum_m \sum_{i,j} \bar{t}^m_{i} t^m_{j} \bar{E}_i E_j$.
Defining the spin variables via the optical phase delays  $\phi_i \in\lbrace0,\pi\rbrace$, so that
 $\sigma_i= exp(\mathrm{i}\phi_i)=E_i$, we obtain (see Appendix A) 
\begin{equation}
I_{T}= - H(\sigma) ;\hspace*{0.2cm} J_{ij}= \sum_{m=1}^M J^{m}_{ij}= \sum_{m=1}^M Re( \bar{t}_{i}^m t_{j}^m ).
\end{equation}
Equation (2) establishes a direct relation between the scattered intensity and the SG energy. Since the TM of a disordered medium is a random full-rank matrix \cite{Popoff2010}, when $M=N$ we find that $J_{ij}$ has uncorrelated random elements with Gaussian distribution $P(J_{ij})$, as for the SG model in Eq. (1). Shaping the binary input phase distribution to maximize the transmitted intensity corresponds to looking for the SG ground state. 
Energy minimization can be performed with any iterative method while the spin system is optically emulated.

\begin{figure*}[t!]
\centering
\vspace*{-0.6cm}
\hspace*{-0.2cm} 
\includegraphics[width=2.1\columnwidth]{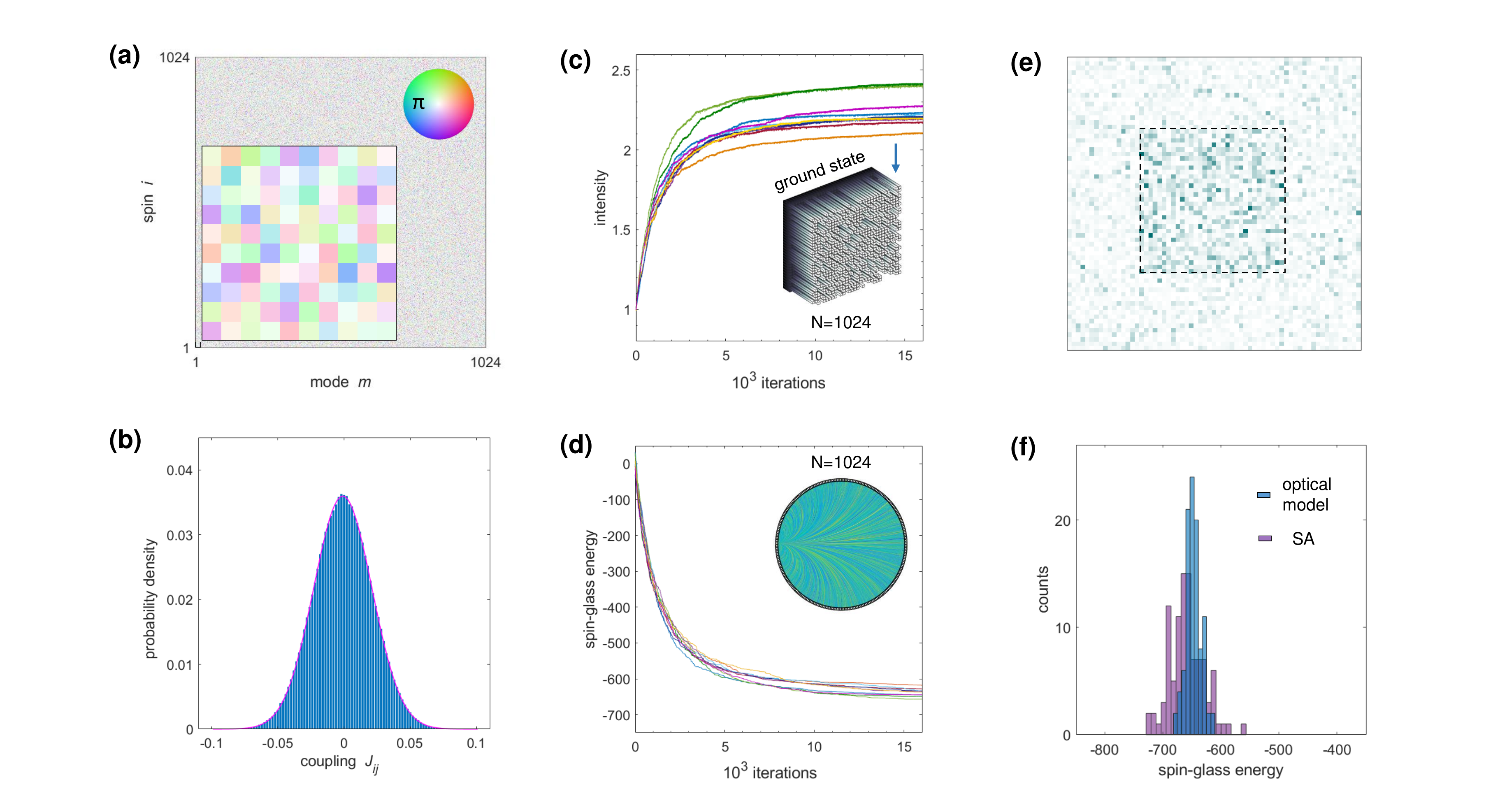} 
\vspace*{-0.7cm}
\caption{ {\bf Design and validation of the optical spin-glass scheme.} (a) TM modeling multiple light scattering between $N=1024$ optical spins 
and $M=N$ detector modes. (b) Coupling's probability distribution with Gaussian fit (line). (c) Total transmitted intensity during the computation 
for different initial conditions and (d) energy of the corresponding SG. Insets in (c-d) show a ground-state phase configuration and its problem graph. 
(e) Final intensity with transmission maximized on the dashed region. (f) Energy histograms of ground states found using the optical SG model and zero-temperature SA.}
\vspace{-0.1cm}
\label{Figure2}
\end{figure*}

\section{Experimental setup and methods}

The optical SG is numerically simulated by forming $N$ pixel blocks from a square mesh (SLM plane).
The initial optical field $E_{I}$ has constant amplitude and its phase is a random configuration of $N$ binary phases, $\phi_i =0, \pi$.
A unitary TM matrix $W$ with random complex numbers is generated. At each iteration, a single spin (phase value $\phi_i$) is randomly selected 
and flipped; the optical field linearly propagates, $E_{T}=W \cdot E_{I}$, and the input phase is updated only if the output total intensity $I_T$ increases.
Numerical evaluation of $I_{T}$ corresponds to a measurement with a $64$-bit sensitivity detector in a noiseless system (Fig. 2).
In general, within this scheme, $\sim10N$ iterations are sufficient for a good convergence.
We normalize the transmitted intensity to the initial transmission, which allows us to compare the result 
with experiments at any input optical power.

The experimental device follows the setup illustrated in Fig. 3(a).
A continuous-wave laser beam at $\lambda= 532$ nm is expanded, polarization controlled, and impinges on 
a reflective liquid-crystal SLM (Meadowlark Optics HSP192-532, $1920\times1152$ pixels) performing phase-only light modulation.
The SLM area is divided into $N$ addressable optical spins by grouping several pixels.
Binary modulated light is projected on the objective back-focal plane (OBJ1, 10x,  NA $=0.1$) and it is focused  
on a strongly scattering medium (a thick diffuser made of teflon, DIFF) with $0.5$mm thickness.
Scattered light is collected by a second objective (OBJ2, 20x,  NA $=0.4$) and the transmitted intensity speckle pattern
is detected by a CCD camera (Basler acA2040-55um,  $2048\times1536$ pixels) with $8$-bit (256 gray-levels) intensity sensitivity on each pixel.
Each camera pixel has a size comparable with the spatial extent of a speckle grain and thus corresponds to an output spatial mode. 
The SLM and CCD have communication bandwidths of $4000$MB/s and $600$MB/s, respectively. 

The ground-state search is conducted sequentially by means of the digital recurrent feedback.
Computation starts from a random configuration of $N$ binary phase blocks (spins) on the SLM. 
The measured intensity distribution determines the feedback signal.
The SNR is approximatively 50, with the main noise sources that are associated to flickering effects of the phase modulator,
power laser fluctuations, and tiny mechanical effects on the scattering medium.
At each machine cycle a batch of spins is randomly selected and flipped;   
the intensity transmitted on $M$ camera pixels is detected and the spin state is updated if the change increases $I_T$.
The batch size is selected as $2.5$\% of the spin number, which ensures that a single change on the SLM is detected over camera noise.
This values determines the maximum achievable accuracy for the experimental ground state. 

\begin{figure*}[t!]
\centering
\vspace*{-0.5cm}
\hspace*{-0.3cm}
\includegraphics[width=2.1\columnwidth]{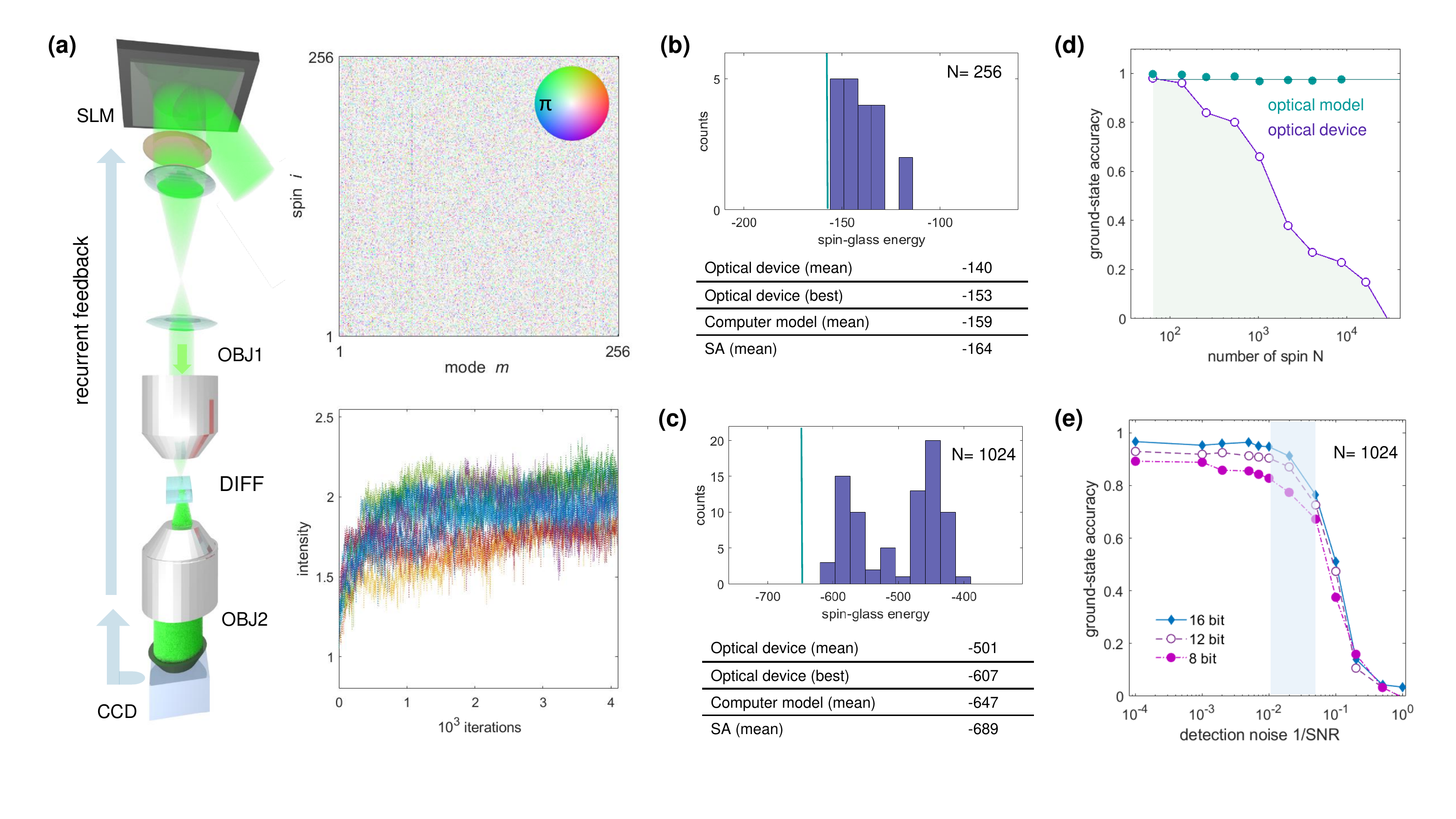} 
\vspace*{-0.9cm}
\caption{{\bf SG computing device.} (a)  Experimental setup, where recurrent feedback from the measured intensity updates the spin configuration on the SLM. Insets show a measured TM and corresponding transmitted intensity during computation for several runs.
(b-c) SG ground-state energy histogram for $N=$256 and $N=$1024. Inset table indicates values for solutions computed optically and with computer algorithms. Vertical lines are average solutions from numerical simulations of the optical SG model. 
(d) Accuracy of the ground-state energy varying the spin number. Experimental results are compared to the scaling behavior of the ideal device.
(e) Modelling of the experimental device: 
ground-state accuracy as a function of optical noise at the detection (SNR, signal to noise ratio) for intensity detectors with different finite precision.
The shaded area indicates the parameter region in which the realized device (8-bit precision) is expected to be operating.
}
\vspace{-0.1cm}
\label{Figure3}
\end{figure*}

\section{RESULTS}

The scheme of the optical SG simulator is numerically validated in Fig. 2, where we model a large-scale device with $N=M=$1024. 
Linear optical propagation through the scattering medium is simulated by the randomly generated TM in Fig.~2(a). According to Eq.~(2),
the TM gives a $J_{ij}$ set following a Gaussian probability density with zero mean and deviation $\bar{J}=1/4N$ (Fig.~2(b) and Supplementary Material \cite{Supplementary}). 
In Fig.~2(c) we show the total intensity, normalized to the initial transmission, that is transmitted on the output modes
during the optimization procedure.
While $I_T$ increases and saturates to a final speckle distribution [Fig.~2(e)], the binary phases on the SLM converge towards a state minimizing the corresponding spin energy [Fig.~2(d)]. The final intensity corresponds to the ground-state energy [Eq.~(2)], apart from a constant factor.
To demonstrate that the method solves the spin problem,
we benchmark the ground-state energies with simulated annealing \cite{Kirkpatrick1983} (SA) at zero temperature on the same random graph.
Results on 100 independent runs are in Fig.~2(f) and indicate that our model operates with an accuracy comparable with a standard robust optimization algorithm. States with lower energy can be found by refining the iterative method by introducing effective temperature variations both in SA
and the optical model (Supplementary Fig.~2 \cite{Supplementary}). 
Simulations on various instances (Supplementary Fig.~1 \cite{Supplementary}) further indicate the effectiveness of our approach in finding heuristic SG solutions.
We find that the low-energy states obtained for different $J_{ij}$ realizations overlaps with the solutions found varying only the initial condition.
Therefore, the SG low-energy space can be sampled by replicating either over the initial spin configuration or realization of the random couplings.

\begin{figure}[t!]
\centering
\vspace*{-0.3cm}
\hspace*{-0.6cm}
\includegraphics[width=1.16\columnwidth]{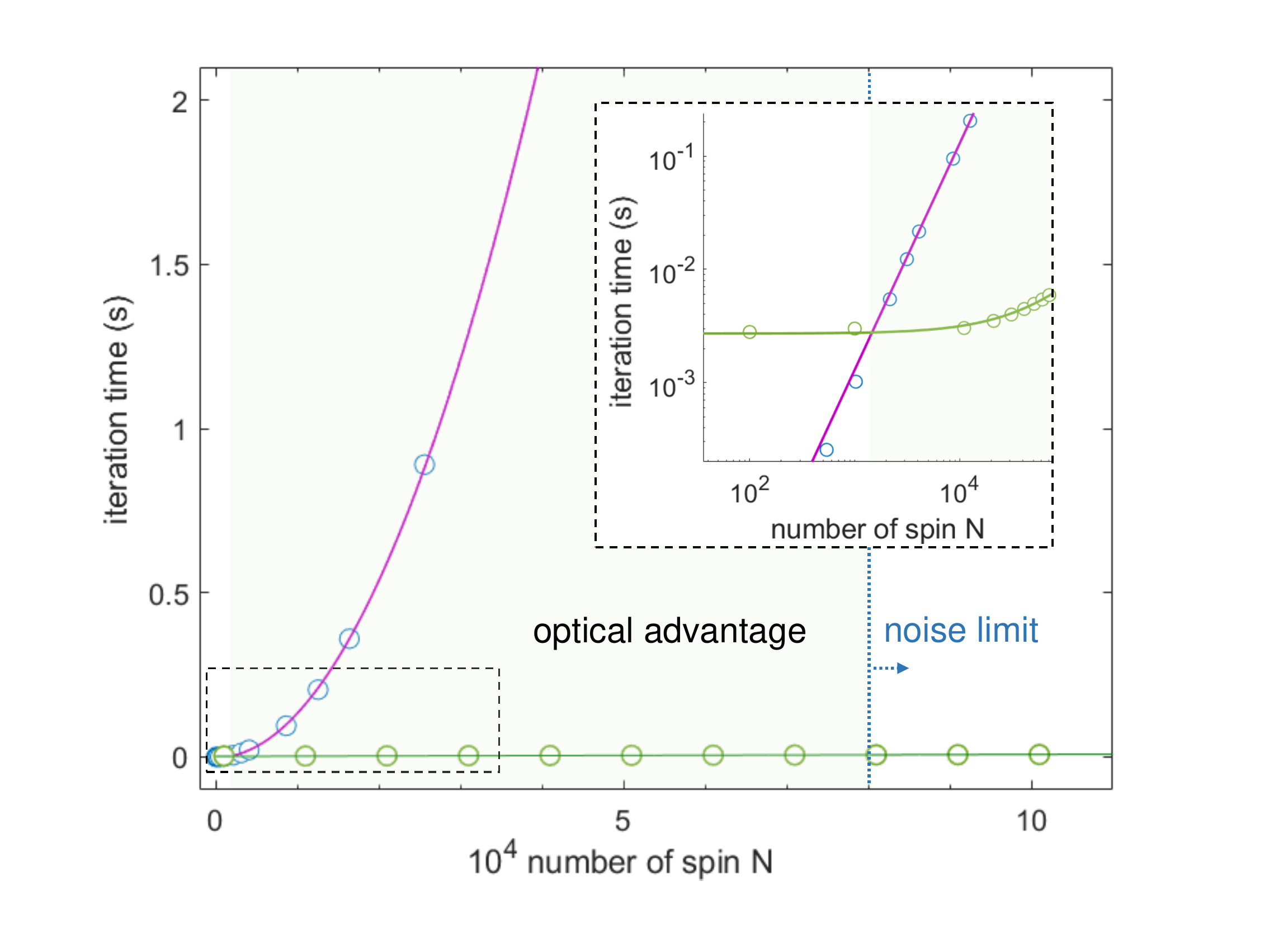} 
\vspace*{-0.6cm}
\caption{ {\bf Optical advantage.} Time for updating the SG configuration (iteration time) versus its size. Values refer to our scheme implemented
on our optical setup (green dots) and on a standard digital computer (blue dots). 
Lines are fit functions showing linear and quadratic scaling of optical and electronic computing.
Inset highlights the setup-dependent crossing point ($N^*=1460$) between the scaling behaviors, which delineate an optical advantage region for large scales (green-shaded area). The vertical dotted line, referred as noise limit, is an hypothetical border to indicate the maximum size solvable with an optical SG simulator given a finite optical noise level.} 
\vspace{-0.1cm}
\label{Figure4}
\end{figure}

\subsection{Experimental implementation}

We realize the optical SG simulator according to the experimental setup in Fig. 3(a). Ising spins are encoded on a laser wavefront by a
phase-only SLM, a volumetric diffuser provides multiple scattering, and camera pixels are the output modes. 
The optical device works in a measurement and feedback scheme. At each machine iteration, we measure the intensity $I_T$ on $M=N$ fixed camera pixels
and update the spins in order to maximize the transmission (see details in Appendix B). 
The setup can make use of any optimization scheme in this operation, i.e.,
it is algorithm-agnostic. After a few thousands of iterations, we get a transmission enhancement (normalized transmitted intensity)
close to the expected value, with variations depending on the the random input condition [inset in Fig. 3(a), $N=$256].
In analogy with the numerical findings, from the measured final intensity we get the ground-state energy for each realization.

Using the optical setting, we performed sets of computations for random SG problems of different sizes, up to more than $10^4$ spins and $10^8$ connections. We quantify the solutions found by analyzing their SG energy in comparison with numerical models. Main results are summarized in Fig.~3(b-c). 
For $N$=256, the Ising machine finds an approximate solution to the NP-hard problem with an accuracy comparable with the optical SG model and SA. 
Optical computing was successful also for large-scale graphs with 1024 nodes, although with lower performance.
For a fixed number of machine iterations (Appendix B), we find the ground-state accuracy decreases as a function of the system size [Fig.~3(d)].
On the contrary, numerical results indicate that the optical SG simulator as modelled in Fig.~2 is able to perform independently of the system size.
The observed behavior is thus a direct consequence of the experimental conditions, i.e., practical non-idealities of the device.
It is important to remark that a similar scaling effect occurs in any other quantum and classical optimizers built in practice,
even at much smaller sizes~\cite{Hamerly2019_2}.

To understand how to improve the computational ability of our proof-of-principle device when the number of spins increases, 
we consider the effect of various experimental factors on the optical SG model.
Specifically, we analyze the dependence of the ground-state accuracy on the detection noise level and
the impact of the finite-precision of the camera at various sizes.
Figure~3(e) shows the results when varying the signal to noise ratio (SNR) for a $N$=1024.
A rapid decrease of the solution accuracy is observed as the noise level exceeds $10^{-2}$,   
which indicates that optical noise partially explains the experimental performance. 
A key role is also played by the finite-precision of the camera. The accuracy improves considerably as we increase the 
detector bit precision, even in presence of noise [Fig. 3(e)].
The effect of the components precision on the computation is even more crucial as the problem size increases (see Supplementary Fig.~4 \cite{Supplementary}).
Additional evidences in Supplementary Material \cite{Supplementary} further indicate that the performance scaling in the realized simulator [Fig.~3(d)] 
is successfully modelled. The overall analysis suggest that,
by concomitantly reducing noise and increasing the precision of the optical readout, the optical simulator can be effective also on large scales. 

From the statistical physics point of view, the operation of the optical SG simulator is limited by a finite effective temperature $T^*$.
We estimate it numerically by using an inverse numerical approach. Exploiting SA, we anneal the SG up to the measured ground-state energy, we let it equilibrate and extract its temperature. For N=4096 we obtain $\beta^*=1/T^* \approx 0.9$. 
Cooling to lower temperatures can be achieved improving the device construction to make it more sensitive and noise tolerant.

A crucial parameter of our spin-glass simulator is the number of output modes. In fact, according to Eq.~(2), 
the transmitted intensity, and thus spin Hamiltonian, depends on the output mode number $M$.
It determines the rank of the interaction matrix and, consequently, the correlations and the distribution of its values.
As reported in Supplementary Fig. 3 \cite{Supplementary}, by increasing $M$ the couplings $J_{ij}$ evolves from sharp peaked distribution to the Gaussian probability density $P(J_{ij})$.
Ising problems in which $M\neq N$, although differ from the considered SK model, can also be particularly interesting for applications. 
For instance, the $M=1$ case directly maps to the number-partitioning problem \cite{Mertens1998}, which is the combinatorial optimization problem 
in which a set of real number must be divided into two subsets differing as little as possible in their weight. The problem is NP-complete and 
represents a typical task encountered in resource allocation~\cite{Lucas2014}. 
For example, in cases where we have to divide a set of assets fairly between two people.
Application of the experimental SG simulator to number partitioning is demonstrated in Supplementary Fig. 5 \cite{Supplementary}. 
The partition solution is optically found with good accuracy. Remarkably, in this case the performance does not degrade with the problem
size~\cite{Fang2020}, and the efficiency is maintained for sets exceeding $10^4$ random numbers.
This underlines that our optical device can be directly applied to specific computing tasks, and can be beneficial in a broad range of applications.
Among these, we mention finding cliques in networks, which is central for understanding social dynamics.

\subsection{Optical advantage}

The key advantage of our optical SG simulator is its possible scalability to sizes intractable with conventional hardware.
In fact, common algorithms require to evaluate Eq.~(1) at each iteration, 
an operation which time and memory consumption grows quadratically with the spin number.
The optical part of our scheme executes such matrix multiplication fully in parallel, independently of the problem size and
feedback algorithm. The scaling advantage is demonstrated in Fig.~4 by measuring the iteration time versus the problem size. 
In contrast to the quadratic scaling of the SG model on a conventional computer, the optical computation time scales only linearly, with a
mild slope depending only on the limited communication bandwidth of the electronic feedback. Therefore, independently of the machine operation frequency,
scaling laws ensure the existence of an optical advantage region at large scales. 
The sensitivity of light modulators and detectors, and more generally, optical noise, rules the maximum size that can be efficiently solved on optical platforms.

\section{CONCLUSION}

In conclusion, we have reported a scalable optical device able to solve random spin problems.
Exploiting spatial light modulation and coherent optical propagation of light, our scheme allows parallel information processing for arbitrary problem 
sizes and without any fabrication constraints. Our setup can be exploited as an optical accelerator for the solution of spin glasses 
with any optimization algorithm \cite{Pierangeli2020_2}.
Given the 8-bit precision of the intensity detector and the effect of experimental noise, the analog spin simulator finds ground states
with energies higher than those obtained with simulated annealing on a 64-bit digital processors. 
However, in principle, the same accuracy can be reached by tuning further the optical setup and the optimization procedure.

The use of a physical medium to encode spin interactions also opens interesting perspectives for programming arbitrary Ising problems, which could be done by selecting various subset of input and output modes \cite{Leedumrongwatthanakun2020, Popoff2019}, or by directly tailoring the transmission matrix using either microfabrication, or a second spatial light modulator \cite{Lvovsky2020}.
Our approach points out a parameters region where optical computing presents a favourable scaling with respect to electronic hardware.
Developments in photonic technology would allow to optically address many NP-hard combinatorial optimizations deep into this region,
where neuromorphic computing can also find its natural application~\cite{Brunner2018, Hamerly2019, Rafayelyan2020}.

\section{APPENDIX A: Spin-glass Hamiltonian in the TM framework} 

The TM models monochromatic transmission through a linear optical system at the mesoscopic level. 
Its complex coefficients $t^m_{i}$ connect amplitude and phase of the optical field between the $m$-th output mode 
and the $i$-th input element, $E_m = \sum_i^N t^m_{i} E_i$, where we adopt superscripted indices only for clarity.
As shown in Ref. \cite{Popoff2010}, the elements $t^m_{i}$ are uncorrelated random complex numbers when a thick disordered medium is placed between a SLM
and a camera, and they can be measured experimentally.
We first consider the intensity on a single output mode, which reads as $I_m= \left| \sum_i t^m_{i} E_i \right|^2 = \sum  \bar{t}^m_j t^m_i \bar{E}_i E_j$.
Defining the spins via the binary phase delays  $\phi_i \in\lbrace0,\pi\rbrace$, so that 
$E_i= exp(\mathrm{i}\phi_i)= \sigma_i$ up to a global phase factor, we get the Ising Hamiltonian $I_m=-H_m=\sum_{ij} J_{ij}^{m} \sigma_i \sigma_j$ with $J_{ij}^ {m}= Re(\bar{t}^m_j t^m_i)$, apart from constant factors.
Pairs of spins with positive (negative) interaction correspond to points of the optical field resulting in constructive (destructive) interference.
In this case, the couplings are correlated, i.e., $rank(J_{ij}^ {m})=1$ (the interaction is specified by only $N$ degrees of freedom).
This case corresponds to a class of Ising problems, known as Mattis SG, that have an exact ground-state solution \cite{Mattis1976}.
In optics such solution corresponds to the optimal wavefront shaping for focusing on a single output mode, 
which, in conditions of negligible noise, gives an enhanced trasmission proportional to $N$ \cite{Gigan2017}.
In combinatorial optimization, such configuration directly maps to number partitioning problem \cite{Fang2020}.
Application of the optical SG simulator for finding specific partitions in a set of random numbers is detailed in Supplementary Material \cite{Supplementary}. 

The interaction matrix and its probability distribution varies considerably when increasing the number of output channels (Supplementary Fig. 3 \cite{Supplementary}).
For $M$ modes, we have $I_{T}= \sum_m I_m = \sum_m \left[ ( \sum_j \bar{t}^m_j  \sigma_j ) (  \sum_i t^m_i \sigma_i ) \right] =
\sum_m \sum_{i,j} \bar{t}^m_{i} t^m_{j} E_i E_j$, which gives the equivalence in Eq.~(2): $I_{T}= - \sum_{ij} J_{ij} \sigma_i \sigma_j$
with $J_{ij}= \sum_{m=1}^M Re( \bar{t}_{i}^m t_{j}^m )$. 
The coupling matrix rank is now $rank(J_{ij})=M$. When $M=N$, we get a full-rank matrix ($N^2$ variables specify the couplings)
describing random uncorrelated spin interactions. The $J_{ij}$ are distributed with a Gaussian density $P(J_{ij})$ (zero mean and deviation $1/4N$), 
as verified in Supplementay Material \cite{Supplementary} for both numerical and experimental data (Supplementary Fig. 1).
Simultaneous maximization of $I_T$ over $N$ output modes is equivalent to minimizing the energy of a SG, with interactions encoded in the TM.

\section{APPENDIX B: Numerical and experimental details}

During the optimization, the SG energy is evaluated by applying Eq. (2) on the optical phase distribution.
At the ground state, such energy is related to the optimized transmitted intensity by a constant factor.
This factor depends only on the spin number. We exploit this property in experiments, where the factor extracted 
from a set of measured TMs at a given size is used also on optical computing runs in which the TM is varied in each realization.
The ground-state accuracy in Fig. 3 is defined 
as $1-[(G-G_{min})/((G+G_{min}))]$, where $G$ and $G_{min}$ are the mean SG energy measured on the device (numerical implementation) 
and its computer model (SA), respectively for purple (green) dots. Error bars indicate one standard deviation over 20 realizations.

To characterize the device in experimental conditions, we introduce various ingredients in the model. 
Finite-precision of the camera is obtained by discretizing the output intensity in $2^n$ levels, $n$ being the number of bits. 
We introduce optical noise at the readout, by adding uncorrelated Gaussian fluctuations to the intensity $I_T$, with the SNR that determines 
the fluctuations amplitude (noise level).
All codes are implemented in MATLAB on an Intel processor with 6 cores running at 4.1 GHz and supported by 16 GB ram.
In Fig.4, iteration times for standard computing refer to this specific CPU.
We note that high-performance computing on dedicated systems can substantially reduce these iteration times; however, 
their quadratic scaling will remain unaltered.
As for simulated annealing (SA), a custom optimized version has been implemented following Ref. \cite{Isakov2015}.
The code exploits various methods including sequential updating, forward energy computation and fast pre-computed random numbers.
It has been benchmarked on standard graphs, including K2000, with results analogous to Ref. \cite{Goto2019} in terms of ground-state energies.

In our scheme, each SG graph corresponds to a measured TM with size $N^2$ \cite{Popoff2010}. The TM is experimentally reconstructed using
Non-negative Matrix Factorization and a phase retrieval algorithm \cite{Boniface2020}.
Slight translations/rotations of the disordered medium result in a different TM.
The optical stability of the scattering medium (approximatively one hour) fixes the physical time for which the interaction matrix remains unaltered.
This factor limits the optimization effectivness over long times. We thus kept the number of iterations in each run constant to 16000
and collect tens of computation varying only the initial condition.
However, faster optical elements can considerably lower the total computation time of our optical SG simulator. 
The optical setup operates at $150$Hz and, according to the employed SLM technology, the iteration time can be reduced up to $1.4$ milliseconds, 
maintaining its linear dependence on the problem size.

\section{Acknowledgements}

\noindent {\bf } D.P. and C.C. acknowledge funding from SAPIExcellence 2019 (SPIM project), QuantERA ERA-NET Co-fund (Grant No. 731473, project QUOMPLEX), PRIN PELM 2017 and H2020 PhoQus project (Grant No. 820392). M.R. and S.G. acknowledge support from H2020 European Research Council (ERC) (Grant No. 724473), DARPA (HR00111890042) and Institut Universitaire de France.
We thank J. Dong and A. Boniface for useful discussions and technical support in the laboratory.

\noindent {\bf Author contributions.} 
D.P. and M.R. carried out experiments, numerical simulations and data analysis.
D.P wrote the paper with contributions from all the authors. C.C. and S.G. co-supervised the project.

\clearpage

\end{document}